\documentclass[twocolumn,aps,prl,showpacs,amsmath,amssymb]{revtex4}
\usepackage{graphicx}% Include figure files
\usepackage{dcolumn}% Align table columns on decimal point
\usepackage{bm}% bold math

\begin{document}

\title{Evolution of the Hall Coefficient and the Peculiar Electronic
Structure of the Cuprate Superconductors}

\author{Yoichi Ando}
 \email{ando@criepi.denken.or.jp}
\author{Y. Kurita}
\altaffiliation{also at Department of Physics, Tokyo University of Science, 
Shinjuku-ku, Tokyo 162-8601, Japan.} 
\author{Seiki Komiya}
\author{S. Ono}
\author{Kouji Segawa}
\affiliation{Central Research Institute of Electric Power Industry, 
Komae, Tokyo 201-8511, Japan.}

\date{\today}

\begin{abstract}

Although the Hall coefficient $R_H$ is an informative transport property
of metals and semiconductors, its meaning in the cuprate superconductors
has been ambiguous because of its unusual characteristics. Here we show
that a systematic study of $R_H$ in La$_{2-x}$Sr$_{x}$CuO$_{4}$ single
crystals over a wide doping range establishes a qualitative
understanding of its peculiar evolution, which turns out to reflect a
two-component nature of the electronic structure caused by an unusual
development of the Fermi surface recently uncovered by photoemission
experiments.

\end{abstract}

%74.25.Fy  Transport properties
%74.72.Dn  La-based cuprates
%74.25.Jb  Electronic structure  

\maketitle

During the past 17 years after the high-$T_c$ superconductivity was
discovered in cuprates, virtually all measurable properties of their
``normal state", the state in the absence of superconductivity, have
been studied to understand the {\it stage} for novel superconductivity.
However, there is yet no established picture for even such basic
properties as the resistivity and the Hall coefficient \cite{Orenstein},
not to mention other more elaborate properties. The Hall coefficient
$R_H$ of conventional metals is independent of temperature and signifies
the Fermi surface (FS) topology and carrier density, but in cuprates
$R_H$ shows strong, sometimes peaked, temperature dependences as well as
a complicated doping dependence. An advance in understanding came when
Chien, Wang and Ong found \cite{Chien} that the cotangent of the Hall
angle, $\cot\Theta_H$ (which is the ratio of the in-plane resistivity
$\rho_{ab}$ to the Hall resistivity $\rho_{H}$), approximately shows a
simple linear-in-$T^2$ behavior, which suggests the existence of a
quasiparticle-relaxation rate that changes as $\sim T^2$. However, while
it appears that the Hall problem in cuprates can be simplified when
analyzed in terms of $\cot\Theta_H$, it was argued by Ong and Anderson
\cite{OngAnderson} that $\cot\Theta_H$ is after all a derived quantity and the
central anomaly resides in the directly measured quantities $\rho_{ab}$
and $R_H$.

In this Letter, we address the notoriously difficult problem of the Hall
effect with the recent knowledge on the physics of {\it lightly-doped
cuprates} and the peculiar evolution of the FS recently elucidated by
the angle-resolved photoemission spectroscopy (ARPES) experiments
\cite{Ino,Yoshida}. We first show that the behavior of $R_H$ and
$\rho_{ab}$ in the lightly-doped cuprates mimics rather well the
behavior of a conventional Fermi liquid, and discuss that this behavior
signifies the physics on the ``Fermi arc", a small portion of the FS
near the Brillouin-zone diagonals. We then discuss that the peculiar
hole-doping dependence and the temperature dependence of $R_H$ reflect a
gradual participation of the ``flatband" near $(\pi,0)$ of the
Brillouin zone, which brings about a sort of two-band nature to the
transport. The measurements of $R_H$ and $\rho_{ab}$ using a standard
six-probe method are done on high-quality single crystals of
La$_{2-x}$Sr$_{x}$CuO$_{4}$ (LSCO) and YBa$_{2}$Cu$_{3}$O$_{y}$ (YBCO),
the details of which have been described elsewhere \cite{mobility}.

In slightly hole-doped LSCO and YBCO, which are usually considered to be
antiferromagnetic insulators, it was demonstrated \cite{mobility} that
the charge transport shows a surprisingly metallic behavior with a hole
mobility comparable to that of optimally-doped superconductors at
moderate temperatures. Detailed ARPES measurements were subsequently
performed on lightly-doped LSCO \cite{Yoshida} and YBCO \cite{Yagi},
which revealed that only patches of FS, called ``Fermi arcs"
\cite{Norman}, are observed at the zone-diagonals, where
quasiparticle-like peaks were detected in harmony with the transport
results. The ARPES results indicate that, for some reason, a significant
fraction of the large FS (that is observed in optimally-doped cuprates
\cite{Shen}) is destroyed and the remaining small portion is responsible
for the metallic transport. Thus, by looking at the transport properties
of the lightly-doped cuprates, one can gain insight into the physics of
the Fermi arc, whose origin is currently under debate \cite{Chakravarty}.

\begin{figure} 
\includegraphics[clip,width=8.5cm]{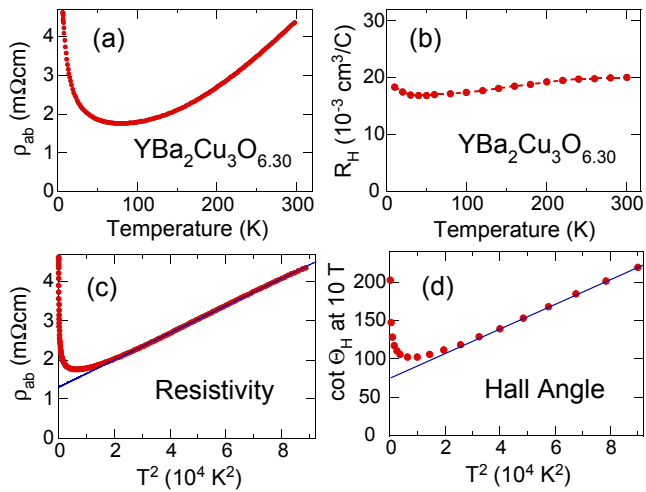}
\caption{Temperature dependences of transport properties of lightly
hole-doped YBa$_{2}$Cu$_{3}$O$_{6.30}$. At this insulating composition,
$\rho_{ab}(T)$ shows positive curvatures (a) and $R_H$ is virtually
$T$-independent at moderate temperatures (b). It turns out that
$\rho_{ab}(T)$ well obeys the $T^2$ law (c), which also holds for 
$\cot \Theta_H$ [$\sim \rho_{ab}/R_H$, (d)], suggesting that a
Fermi-liquid-like $T^{2}$ scattering rate governs the transport on the
Fermi arcs.} 
\end{figure}

Figs. 1(a) and 1(b) show an example of the behavior of $\rho_{ab}(T)$
and $R_H(T)$ in a lightly-doped cuprate: here the data are for YBCO with
$y$ = 6.30 (hole doping of about 3\% per Cu \cite{mobility}), which is
an antiferromagnet with the N\'eel temperature of 230 K \cite{Ando}. As
we have reported previously \cite{mobility,MOS}, $R_H$ is virtually
$T$-independent (as in conventional metals) at moderate temperatures in
the lightly-doped samples; as a result, $\rho_{ab}$ and $\cot \Theta_H$
have the same $T$ dependence [see Figs. 1(c) and 1(d)], which,
intriguingly, is most consistent with $\sim T^2$ and not with $\sim T$.
This implies that the relaxation rate of the ``quasiparticles" on the
Fermi arc changes as $\sim T^{2}$, which incidentally is the same as
the behavior of conventional Fermi liquids. Note that the
low-temperature upturn in $\rho_{ab}$ and $R_H$ is due to localization
effects \cite{anisotropy}, which just obscure the intrinsic
low-temperature behavior of the system.

We found that the $T$ dependence of $\rho_{ab}$ is consistent with 
$\sim T^2$ not only in lightly-doped YBCO but also in lightly-doped LSCO, 
as shown in Fig. 2(a) for $x$ =
0.02; however, it should be noted that the temperature range for the
$T^2$ law is a bit narrow [Fig. 2(a) shows the data up to 250 K] and
$\rho_{ab}(T)$ deviates downwardly from the $T^2$ dependence at high
temperatures. As is shown for $x$ = 0.08 [Fig. 2(b)], the deviation
tends to start from lower temperature at larger $x$. One possible way to
interpret this deviation is to ascribe it to an increase in the density
of states at the Fermi energy $E_F$ with increasing $T$, that happens,
for example, when a gap is filled in with $T$. In fact, the band
structure of LSCO elucidated by ARPES \cite{Ino,Yoshida} suggests such a
possibility: In lightly-doped LSCO a ``flatband" [located near
$(\pi,0)$ of the Brillouin zone] lies below $E_F$ and this band
gradually moves up to $E_F$ with increasing doping; therefore, if
thermal activation causes some holes to reside on the ``flatband" and
to contribute to the conductivity, the high-temperature deviation from
the $T^2$ behavior and its doping dependence can be understood, at least
qualitatively.  

The systematics of $R_H(T)$ and $\cot \Theta_H(T)$ of LSCO for a very
wide range of doping ($x$ = 0.02 -- 0.25) is shown in Fig. 3. Let us
first compare the robustness of the $T^2$ behavior in $\cot \Theta_H$ to
that in $\rho_{ab}$: One can see in Fig. 3(b) that $\cot \Theta_H$ for
both $x$ = 0.02 and 0.08, where the behavior of $\rho_{ab}$ was
discussed, shows no high-temperature deviation from the $T^2$ behavior
up to 300 K. This contrast between $\rho_{ab}(T)$ and $\cot \Theta_H(T)$
regarding the robustness of the $T^2$ law is qualitatively
understandable if both $\rho_{ab}$ and $R_H$ reflect a change in the
effective carrier density $n_{\rm eff}$ due to the temperature-dependent
participation of the flatband, since such a change tends to be
cancelled in $\cot \Theta_H$; remember, $H\cot \Theta_H$ is equal to the
inverse Hall mobility and, thus, would normally be free from a change in
the carrier density. This observation suggests that the relative
simplicity in the behavior of $\cot \Theta_H$ comes from its lack of a
direct dependence on $n_{\rm eff}$, while both $\rho_{ab}$ and $R_H$ depend
directly on $n_{\rm eff}$.

\begin{figure}
\includegraphics[clip,width=8.5cm]{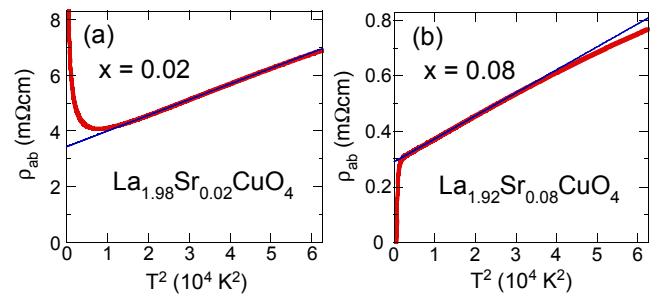}
\caption{Validity of the $T^2$ law in $\rho_{ab}(T)$ of LSCO single
crystals. At a lightly-doped composition $x$ = 0.02 (a), the 
$T^2$ law (shown by a solid line) holds for the 
temperature range of 130 -- 230 K, while at $x$ = 0.08 (b) the range 
is 60 -- 160 K.}
\end{figure}

Thus, our data are most consistent with an emerging picture that a
Fermi-liquid-like transport results from the ``quasiparticles" on the
Fermi arcs in lightly hole-doped cuprates, and the rest of the FS starts
to contribute to the transport at higher doping and/or temperature. It
should be noted, however, that there cannot be a real ``Fermi liquid" on
the Fermi arcs, because the large magnitude of $\rho_{ab}$ in the
lightly-doped cuprates would indicate that the mean free path of the
electrons at $E_F$ is {\it shorter} than their de Broglie wavelength
\cite{mobility}, which is impossible in a Fermi liquid; it was proposed
that some self-organized inhomogeneity in the real space (such as charge
stripes \cite{Carlson}) would offer a resolution to this apparent puzzle
\cite{mobility}. Also note that the patchy Fermi arcs which do not
enclose a well-defined area cannot host a Fermi liquid. Therefore, the
lightly-doped cuprates are peculiar in that what appears to be
conventional Fermi-liquid-like behavior characterizes the system which
cannot really be a Fermi liquid.

\begin{figure}
\includegraphics[clip,width=8.5cm]{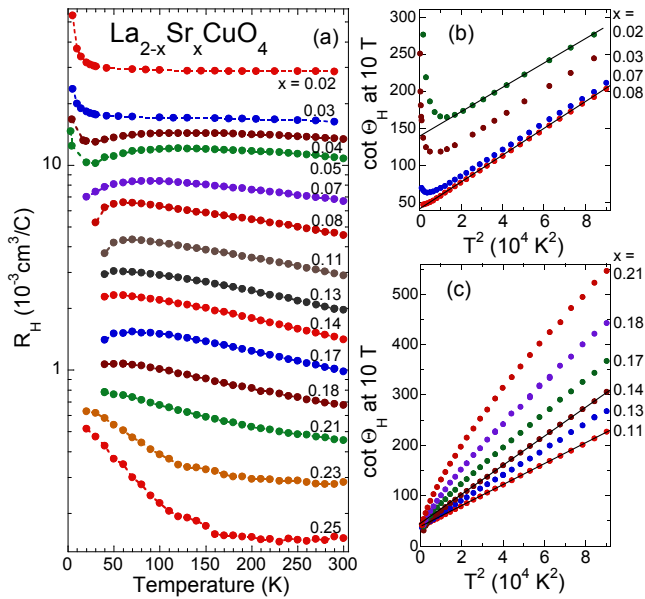}
\caption{Hall response in LSCO single crystals for a wide range of
doping. (a) Variation of the $T$-dependence of $R_H$ for 
$x$ = 0.02 -- 0.25, all measured on high-quality single crystals; 
the $x$ values are determined by the inductively-coupled-plasma
atomic-emission-spectroscopy analyses and are accurate within $\pm$5\%.
(b,c) Plots of $\cot \Theta_H$ vs. $T^2$ for representative 
$x$ values; for selected data, solid lines emphasize the $T^2$ law 
in $\cot \Theta_H(T)$, which holds well for $0.02 \le x \le 0.14$.}
\end{figure}

In passing, we note that in Fig. 3(c) the $T^2$ law of $\cot \Theta_H$
holds very well up to optimum doping ($x \simeq 0.16$), but gradually
breaks down when samples are overdoped; this observation confirms the
trend previously noted by Hwang {\it et al.} \cite{Hwang} for
polycrystalline samples. It is tempting to associate this change in the
behavior of $\cot \Theta_H$ to the putative quantum phase transition
\cite{Orenstein} at optimum doping in LSCO, for which other transport
properties also give indirect evidence \cite{Boebinger,Sun}.
Interestingly, similar change in the power-law behavior of $\cot
\Theta_H$ has been observed to occur above $p \simeq 1/8$ ($p$ is the
hole doping per Cu) in Bi$_2$Sr$_{2-x}$La$_x$CuO$_{6+\delta}$
\cite{Murayama}, where the transition between ``metal" and ``insulator"
under 60 T was found to lie also at $p \simeq 1/8$ \cite{Ono}. It has
been discussed that these crossover phenomena and some of the
non-universalities may be attributed to competing orders
\cite{Orenstein,Yeh}; in this regard, it is useful to mention that the
asymmetries between hole-doped and electron-doped cuprates may also be
due to some sort of competing orders \cite{Yeh}.

\begin{figure}
\includegraphics[clip,width=8.0cm]{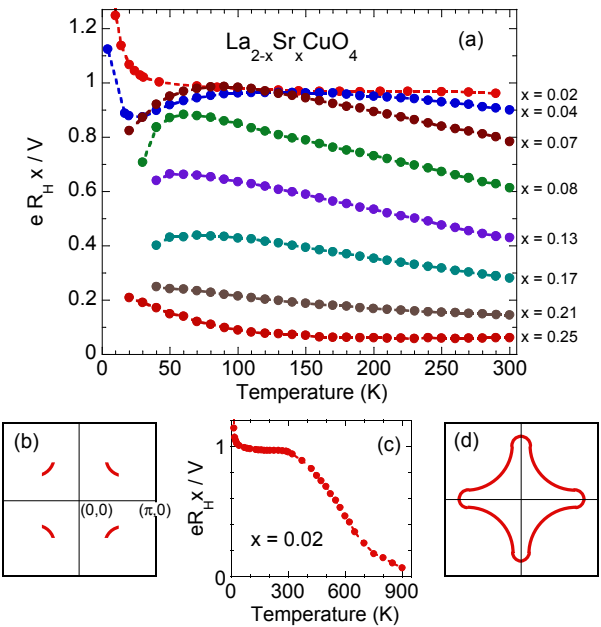}
\caption{Evolution of $R_H(T)$ and the Fermi surface in LSCO.
(a) Temperature dependences of $e R_H x/V$, which gives the
ratio of measured Hall coefficient to that expected from nominal 
hole density, $x/V$, for representative $x$ values; 
higher temperature data up to 900 K are shown for $x$ = 0.02 (c). 
Schematic pictures of the Fermi arcs (b) and the ``electron-like" 
closed FS (d) are also shown. The $T$-independent behavior of 
$R_H$ observed at moderate temperatures for the two extremes, $x$ = 0.02 
and 0.25, measures the carrier density on the Fermi arcs and on the 
closed FS, respectively, and can be rather classically interpreted.}
\end{figure}

Next we discuss the evolution of $R_H$ with doping in more detail. It is
instructive to plot $e R_H x/V$ vs. $T$ for various dopings ($e$ is
electron charge and $V$ is unit volume per Cu); note that $e R_H x/V$
should be 1 if $R_H$ simply signifies the nominal hole density, $x/V$.
Fig. 4(a) shows such a plot for LSCO at representative dopings. One can
see that $e R_H x/V$ is actually 1 at around 100 K for $x$ = 0.02 --
0.07, but it tends to decrease from 1 with increasing temperature and
this decrease starts from lower temperature as $x$ becomes larger. This
behavior suggests that on the Fermi arcs [Fig. 4(b)], which appear to
govern the transport in these lightly-doped samples at $\sim$100 K, the
carrier density is exactly equal to the nominal density of doped holes.
As $x$ is increased above 0.07, $e R_H x/V$ becomes smaller and never
reaches 1; this trend is most likely related to the evolution of the
``flatband" that we already mentioned, because this band touches $E_F$
for $x \agt 0.10$ \cite{Ino}. Therefore, it appears that the magnitude
of $R_H$ actually reflects the effective carrier density which becomes
larger than the nominal hole density when the flatband starts to
participate in the physics near $E_F$ in addition to the Fermi arcs as
doping and/or temperature is increased. 

To further investigate the effect of temperature, we have measured $R_H$
for $x$ = 0.02 up to 900 K, and the resulting data [Fig. 4(c)] turn out
to be surprisingly informative: they suggest that even at such low
doping as $x$ = 0.02 the flatband starts to participate in the Hall
response above 300 K, which implies that the ${\bf k}$-space patchiness
melts away with thermal fluctuations and a full Fermi surface is
eventually restored. Also informative is the behavior of $R_H$ in our
most overdoped sample, $x$ = 0.25, where $R_H$ is essentially $T$
independent down to $\sim$150 K. For overdoped LSCO, ARPES found
\cite{Ino} that the FS is ``electron like" with the shape depicted in
Fig. 4(d); since this FS contains both positive- and negative-curvature
parts, even within the conventional Boltzmann transport theory the value
of $R_H$ is determined by a rather complicated balance between the
contributions from the two parts \cite{Ong}. Remember that the positive
(negative) curvature of the FS results in hole (electron) like Hall
response. Therefore, one can interpret that the $T$-independent part of
$R_H$ for $x$ = 0.25 is actually determined by the shape of the FS in a
classical way, while the low-temperature upturn can be due to a
development of the pseudogap \cite{Hwang} and/or some anisotropy in the
relaxation rate on the FS \cite{Ong}. This interpretation also naturally
explains the sign change in the high-temperature value of $R_H$ in
further overdoped samples \cite{Hwang}, for which ARPES showed that the
positive-curvature part of the FS is gradually diminished \cite{Ino},
making electron-like Hall response to become dominant. Note that the
precise $T$-dependence of $R_H$ for a given $x$ is likely to be governed
both by the participation of the flatband and by the anisotropic
relaxation rate in a complicated way, which is probably the reason why
the Hall effect in the cuprates has been so difficult to be understood.

\begin{figure}
\includegraphics[clip,width=8.5cm]{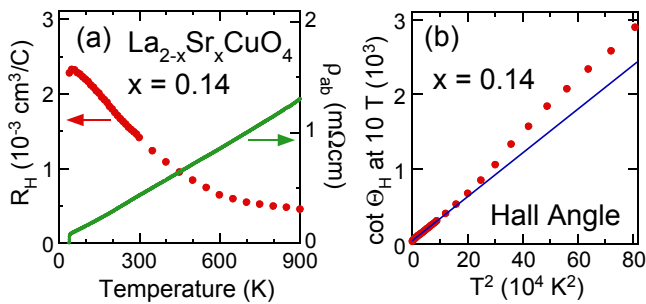}
\caption{(a) Temperature dependences of $\rho_{ab}$ and $R_H$ for LSCO at 
$x$ = 0.14 up to 900 K. (b) $T^2$ plot of $\cot \Theta_H$, which demonstrates 
that the $T^2$ law breaks down above $\sim$400 K.}
\end{figure}

One may notice that a natural extension of the present argument would
be that the $T$-linear resistivity usually observed near optimum
doping may not necessarily be a sign of a $T$-linear relaxation rate, 
because $n_{\rm eff}$ may be changing with $T$. 
To address this
issue, we show data for $\rho_{ab}$, $R_H$, and $\cot \Theta_H$ of LSCO
at $x$ = 0.14 up to 900 K in Fig. 5. One can see that
$R_H$ tends to saturate at high $T$ (as it happens for $x$ = 0.25)
and the $T^2$ law in $\cot \Theta_H$ breaks down above $\sim$400 K, 
both of which suggest that there is an electronic crossover with 
increasing temperature (this can be the pseudogap crossover, see 
Ref. \cite{Hwang}). Therefore, while the $T$-linear resistivity is 
indeed observed up to 900 K [Fig. 5(a)], it is not very 
likely that $n_{\rm eff}$ stays constant and 
the relaxation rate is $T$-linear up to 900 K.
It is worthwhile to mention that the question of whether $n_{\rm eff}$
is changing with temperature or just the relaxation rate is changing is
similar to the question regarding the interpretation of the optical
conductivity $\sigma_{1}(\omega)$ of cuprates \cite{optics}; namely, the
peculiar $\omega$ dependence of $\sigma_{1}$ can be interpreted either
with a two-component model (simple Drude + mid-infrared resonance) or
with an extended Drude model (where the relaxation rate changes with
$\omega$). The former allows more electrons to respond to the electric
field at higher energy, which bears similarities to our ``two-band"
picture (``Fermi arc" + ``flatband") for the dc transport.

Thus, although the detailed situation at optimum doping needs further
clarification, the present results demonstrate that the physics of the
normal state is phenomenologically simple in the two extremes,
lightly-doped regime and the heavily-overdoped regime; in particular,
$R_H$ has a classical meaning in these two extremes at moderate
temperatures. The evolution from one to the other is a transition in the
${\bf k}$ space from the ``Fermi arc" to a closed Fermi surface
(patchiness to continuity), upon which the ``flat-band" appears to
provide additional carriers and play a crucial role. Interestingly,
increasing temperature also causes the same transition from patchiness
to continuity. If the peculiar transport properties in the lightly-doped
cuprates are inherently related to a self-organized inhomogeneity in the
real space \cite{Kivelson} as some resent observations seem to indicate
\cite{mobility,Ando,anisotropy,MR}, the patchiness in the ${\bf k}$
space is likely to be the other side of the same coin. Therefore, a key
to understanding the cuprates probably lies in the dual transition from
patchiness to continuity in the ${\bf k}$ space and from inhomogeneity
to homogeneity in the real space.

We thank D. N. Basov, A. Fujimori, S. A. Kivelson, A. N. Lavrov, and 
N. Nagaosa for helpful discussions.

\end{document}